# High Performance Metal-Insulator-Graphene Diodes for Radio Frequency Power Detection Application

Mehrdad Shaygan[1,*], Zhenxing Wang[1,*], Mohamed Saeed Elsayed[2], Martin Otto[1], Giuseppe Iannaccone[3], Ahmed Hamed Ghareeb[2], Gianluca Fiori[3], Renato Negra[2], Daniel Neumaier[1]

[1]Advanced Microelectronic Center Aachen (AMICA), AMO GmbH, Otto-Blumenthal-Str. 25, 52074 Aachen, Germany

[2]Chair of High Frequency Electronics, RWTH Aachen University, Templergraben 55, 52056 Aachen, Germany

[3]Dipartimento di Ingegneria dell'Informazione, Università di Pisa, Via Caruso 16, Pisa 56122, Italy

[*]Author to whom correspondence should be addressed. Email: shaygan@amo.de (MS), wang@amo.de (ZW).

**Abstract**

Vertical metal-insulator-graphene (MIG) diodes for radio frequency (RF) power detection are realized using a scalable approach based on graphene grown by chemical vapor deposition and $TiO_2$ as barrier material. The temperature dependent current flow through the diode can be described by thermionic emission theory taking into account a bias induced barrier lowering at the graphene $TiO_2$ interface. The diodes show excellent figures of merit for static operation, including high on-current density of up to 28 A/cm$^2$, high asymmetry of up to 520, strong maximum nonlinearity of up to 15, and large maximum responsivity of up to 26 V$^{-1}$, outperforming state-of-the-art metal-insulator-metal and MIG diodes. RF power detection based on MIG diodes is demonstrated, showing a responsivity of 2.8 V/W at 2.4 GHz and 1.1 V/W at 49.4 GHz.

**Introduction**

Diodes are key components for radio frequency (RF) electronic devices. They are typically used as non-linear devices for RF signal detection in RF receivers, for frequency multiplication and signal up/down-conversion, or for RF energy harvesting applications. [1-5] The performance of a diode is defined by the diode's non-linear characteristic and the maximum operation speed, which is usually given by the transit time of charge carriers or the resistance-capacitance (RC) product. Currently semiconductor p-n junction based diodes and Schottky diodes are dominating RF applications because of their excellent non-linear characteristic, well controlled processing technology and high RF bandwidth. [3-5] However, in recent years, there was a growing interest in metal-insulator-metal (MIM) diodes for future RF applications, since MIM diodes i) are expected to operate at higher frequencies because of the very low series resistance and ii) can be fabricated with thin-film processing techniques, mandatory for applications on alternative substrates as glass or plastic, as needed in e.g. flat plane display applications. [6-11] However, their performance in terms of nonlinearity and static performance is still behind traditional semiconductor based diodes.

Recently a metal-insulator-graphene (MIG) diode has been reported based on mechanically exfoliated graphene, demonstrating higher static performance in terms of nonlinearity and asymmetry compared to state-of-the-art MIM diodes [12] because of a bias induced barrier height modulation at the graphene-insulator interface. However, operation at RF has not yet been demonstrated for MIG diodes and it remains unclear if such high performance can also be achieved by using large scale graphene.

In this work, we report on the scalable fabrication and characterization of MIG diodes based on large scale chemical vapor deposition (CVD) grown graphene and a $TiO_2$ barrier formed by atomic layer deposition (ALD). The fabricated MIG diodes show excellent figures of merit (FOMs), including on-current density, asymmetry, nonlinearity and responsivity outperforming the state-of-the-art for MIM and MIG diodes. RF power detection is demonstrated for the first time for a MIG diode and a detection responsivity of 2.8 V/W at 2.4 GHz and 1.1 V/W at 49.4 GHz is observed.

**Results and Discussion**

A schematic cross section of the MIG diode is shown in Fig. 1a, comprising an embedded metal electrode, a $TiO_2$ barrier layer and graphene contacted by a metal electrode on top. The choice of such a geometry has been motivated by the fact that high quality $TiO_2$ barrier layer can be grown directly on the metal using plasma assisted ALD, while avoiding the challenging and less reproducible deposition of thin dielectric layers on top of graphene. The MIG diodes were fabricated on a high-resistivity Si substrate (resistivity ~ 5 kΩ·cm) with 1 µm thermal $SiO_2$ in order to minimize the parasitic capacitance through the substrate. The entire fabrication process was performed using optical contact lithography. As first step 200 nm deep trenches were etched into $SiO_2$ by reactive ion etching (RIE) for the deposition of the embedded metal electrodes. With the same resist mask, the trenches were then filled with a stack of 180 nm Al and 20 nm Ti using e-beam evaporation, followed by lift-off. A layer of 6 nm $TiO_2$ was subsequently deposited by ALD at 300°C using an oxygen plasma process with titanium tetrachloride ($TiCl_4$) as precursor. Vias through the $TiO_2$ layer were opened by sputtering with Ar plasma and subsequently sealed with 20 nm Ni without breaking the vacuum. Commercially available graphene grown by CVD on copper foil (Graphenea SE) was transferred onto the sample using PMMA as a supporting layer [13]. After patterning the graphene by means of oxygen plasma, metal contacts to the graphene were fabricated by sputter deposition and lift-off, with a 20 nm Ni and 100 nm Al metal stack. The I-V characterization of the diodes is done at room temperature in nitrogen. Figs. 1b and 1c show optical images of the fabricated device array.

The current-voltage (*I-V*) characteristic of one MIG diode is shown in Fig. 2a with a clear non-linear and asymmetric diode characteristic, with the forward direction at a positive voltage applied to the Ti-electrode, while the graphene electrode was kept grounded. The current density reaches values of 3.6 A/cm$^2$ and 28 A/cm$^2$ at 1 V and 1.2 V bias, respectively (the area of the diode is 140 μm$^2$), much higher compared to that of the state-of-the-art MIM or MIG diodes (see Table 1). The ideality factor $\eta$ of the MIG diode is 1.7, extracted according to the following relationship [14]

$$\ln I \propto \frac{e}{\eta k_\text{B} T} V \qquad (1)$$

where $e$ is the elementary charge, $k_\text{B}$ is the Boltzmann constant and $T$ is the temperature. The *I-V* characteristic of the diode is plotted in Fig. 2b for temperatures of 295 K, 313 K, 333 K, 353 K and 373 K. While the overall characteristic of the diode in terms of asymmetry and nonlinearity is not significantly affected by the temperature, the absolute current density shows a strong temperature dependency, i.e. a clear signature of thermal activated transport across the barrier, as for thermionic emission.

In order to provide a physical understanding of the mechanism that dominates the operation of the MIG diodes, we have performed simulations based on a top-of-the-barrier model as in [15], computing the electrostatic potential distribution along the MIG structure. The band edge profiles for different applied biases (-1, 0 and 1 V) are shown in Fig. 2c. As can be seen, the barrier seen by electrons emitted from graphene (i.e. the difference between the electrochemical potential of graphene and the top of the conduction band at the graphene/TiO$_2$ interface) varies for the different applied bias voltage. For the positive bias the barrier height for electron transport from graphene to metal Ti ($\Phi_\text{G}$) dominates the current flow, while for the negative bias the barrier height for electron transport from metal Ti to graphene ($\Phi_\text{M}$) dominates. The dominating barrier height $\Phi_\text{G}$ for the current flow is plotted in Fig. 2d. The barrier height $\Phi_\text{G}$ decreases with increasing voltage, i.e. the bias induced barrier lowering, which explains the high on-current for the diode at forward bias. All of the above explains the experimentally observed asymmetry in the *I-V* characteristic. We note that for MIG diodes the ideality factor depends on the bias dependent barrier height from the graphene side. The ideality factor is expected to become 1 for $\Phi_G \propto V$, which is only the case if the geometric capacitance is significantly larger than the quantum capacitance of graphene. In our case an ideality factor of 2 is expected according to the simulations, which is in good agreement to the experimental value of 1.7.

In order to evaluate the MIG diode for the application in rectifiers or power detectors, the typical FOMs are extracted from the *I-V* characteristics, [7, 9, 11] in particular, the asymmetry of a diode defined as

$$f_{\text{ASYM}} = \left| \frac{J_F}{J_R} \right|, \tag{2}$$

where $J_F$ and $J_R$ are the current density for forward and reverse bias, respectively, the nonlinearity as

$$f_{\text{NL}} = \frac{dJ}{dV} \Big/ \frac{J}{V}, \tag{3}$$

where *J* is the current density of the diode, and the responsivity as

$$f_{\text{RES}} = \frac{d^2J}{dV^2} \Big/ \frac{dJ}{dV}. \tag{4}$$

As shown in Fig. 2a, the *I-V* characteristics exhibit a clear asymmetry, with $f_{\text{ASYM}}$ increasing for increasing biases, reaching a maximum value of 270 at 1.1 V bias, as shown in Fig. 3a. The extracted $f_{\text{NL}}$ is a measure of the deviation of the diode performance from the linear behavior. As shown in Fig. 3b, the nonlinearity increases with the bias voltage, reaching a maximum of 11 at 0.7 V. The maximum responsivity of 24 V$^{-1}$ is obtained at 0.2 V (see figure 3c).

We have also performed a statistical investigation of the extracted FOMs on a total of 11 devices from one single chip, including the one shown in Figs. 2 and 3. The extracted FOMs are shown in the histograms in Fig. 4. For comparison, the average and maximum values for the different FOMs are listed in Table I, and compared to state-of-the-art MIM and MIG diodes. It is clearly demonstrated that even the obtained average values for on-current density, nonlinearity, and responsivity outperform all the MIM and MIG diodes.

To demonstrate power detection at RF frequencies, we used the setup illustrated in Fig. 5a. A Rohde & Schwarz ZVA-50 vector network analyzer was used as a calibrated power source with known incident power for the frequency range from 2 GHz to 50 GHz. The incident power is coupled with a DC bias to the diode using a bias tee. Impedance matching is used to match the input of the diode to the incident RF signal. In order to measure the DC signal resulting from the rectification of the RF signal, we use a low pass filter at the output of the diode. The RF power source and the DC meter as well as the data acquisition are communicated via GPIB cables to the controlling PC. The power detection characterization is done at room temperature in ambient. The diode is biased at 0.5 V, and the measured DC output signal versus input RF power for different signal frequencies are plotted in Fig. 5b. The frequency points are chosen based on the measured reflection coefficient from S-parameters ($S_{11}$) during power source calibration, in order to ensure a proper matching at the input interface. The RF responsivity of the detector can be calculated based on the slope of the plot. For 2.4 GHz input frequency, the responsivity reaches 2.8 V/W. For higher frequencies, the responsivity drops to 2.2 V/W at 30.8 GHz and 1.1 V/W at 49.4 GHz input frequency without de-embedding substrate losses. Improvements can be expected if the input matching together with the resistive load can be integrated on chip.

**Conclusions**

Scalable fabrication of MIG diodes is demonstrated by using CVD grown graphene and $TiO_2$ as a barrier deposited by plasma ALD. The MIG diodes show excellent static FOMs outperforming state-of-the-art MIMs and MIGs diodes, i.e., high on-current density up to 7.5 A/cm$^2$ for 1 V bias, high maximum asymmetry up to 520, strong maximum nonlinearity up to 15, and large maximum responsivity up to 26 V$^{-1}$. By means of temperature dependent characterization and numerical simulations, we have demonstrated that the thermionic emission of electrons is the dominant mechanism for conduction. RF power detection based on the MIG diode is demonstrated for input frequencies from 2.4 GHz up to 49.4 GHz, showing a linear detection responsivity of 2.8 V/W at 2.4 GHz and 1.1 V/W at 49.4 GHz.

**Acknowledgements**

This work was financially supported by the European Commission under the projects Graphene Flagship (contract no. 696656), SPINOGRAPH (contract no. 607904) and by the German Science Foundation (DFG) within the priority program 1796 FFlexCom Project "GLECS" (contract no. NE1633/3).

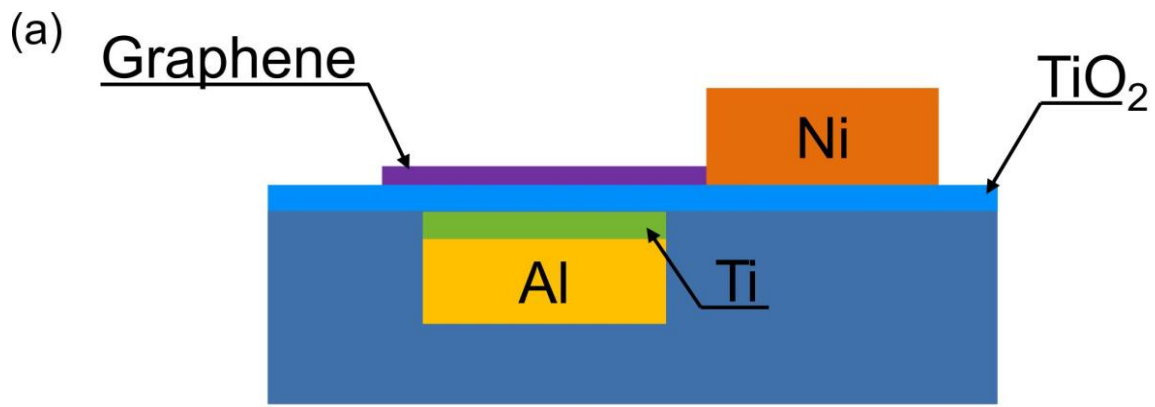

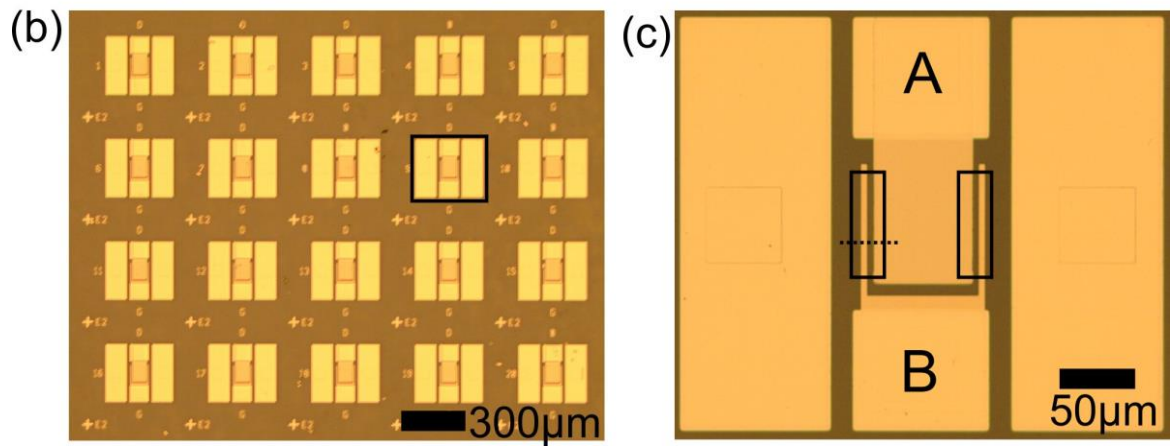

Figure 1. The schematic and optical images of the MIG diodes. (a) Schematic side view of the diode. (b) Microscope image of the device array. Scale bar is 300 μm. (c) Microscope image of the device in the black box of (b), with the layout for high frequency measurement. Scale bar is 50 μm. The contacting electrode to graphene is marked with A and the embedded electrode is marked with B. Graphene areas are marked with the black boxes. The dashed line shows the position where the schematic cross section view of (a) is taken.

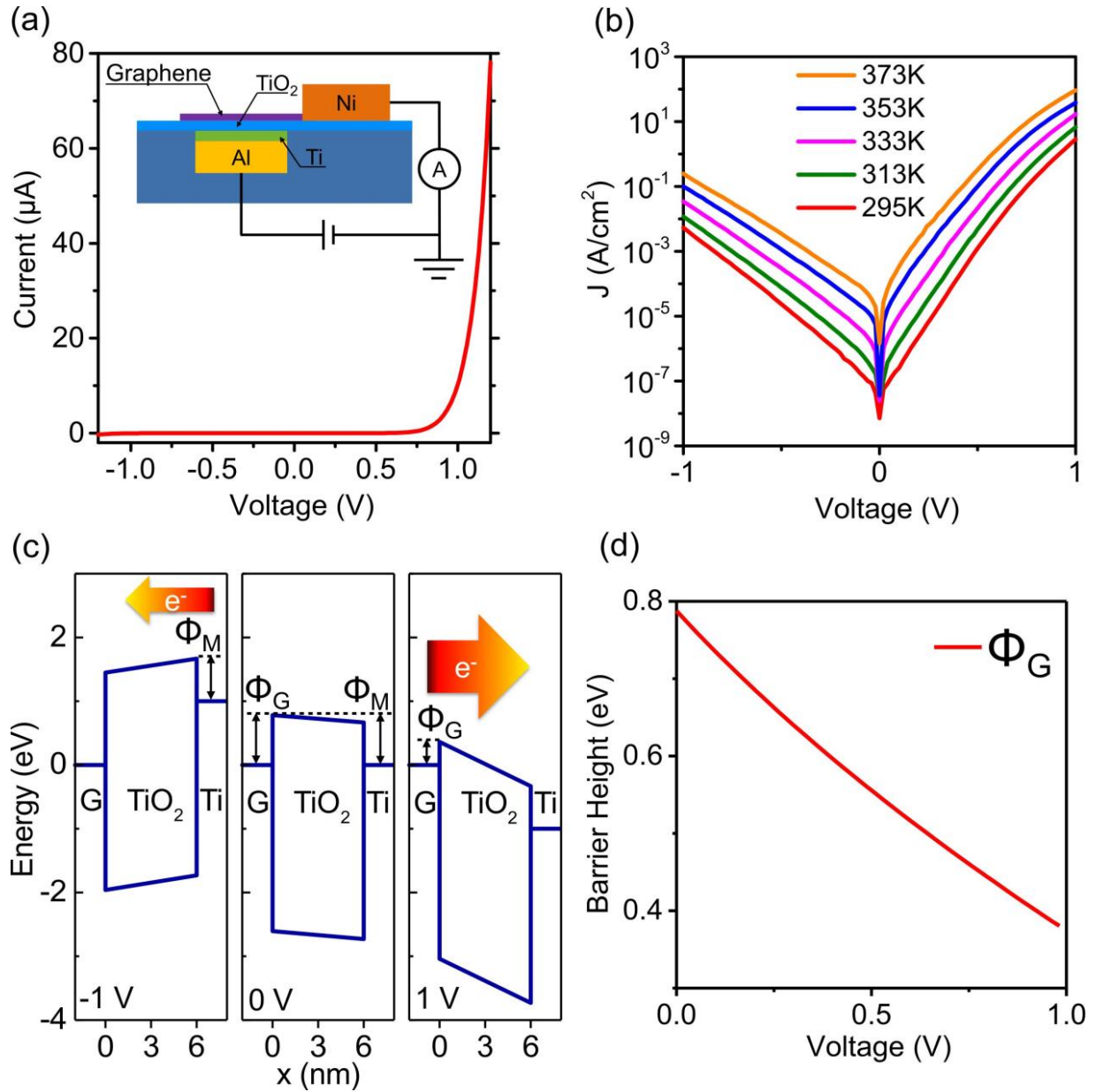

Figure 2. DC characterization of the MIG diode and corresponding simulations. (a) *I-V* characteristics of the diode. Inset: the measurement setup. (b) Current density-voltage (*J-V*) curve of the diode for different temperatures on a semi-logarithmic scale. (c) Band diagrams of the MIG diode at different biases (-1, 0, and 1 V from left to right respectively), which are computed through numerical simulations. (d) The simulated dominating barrier height for electron transport from graphene to metal Ti ($\Phi_G$) at different biases.

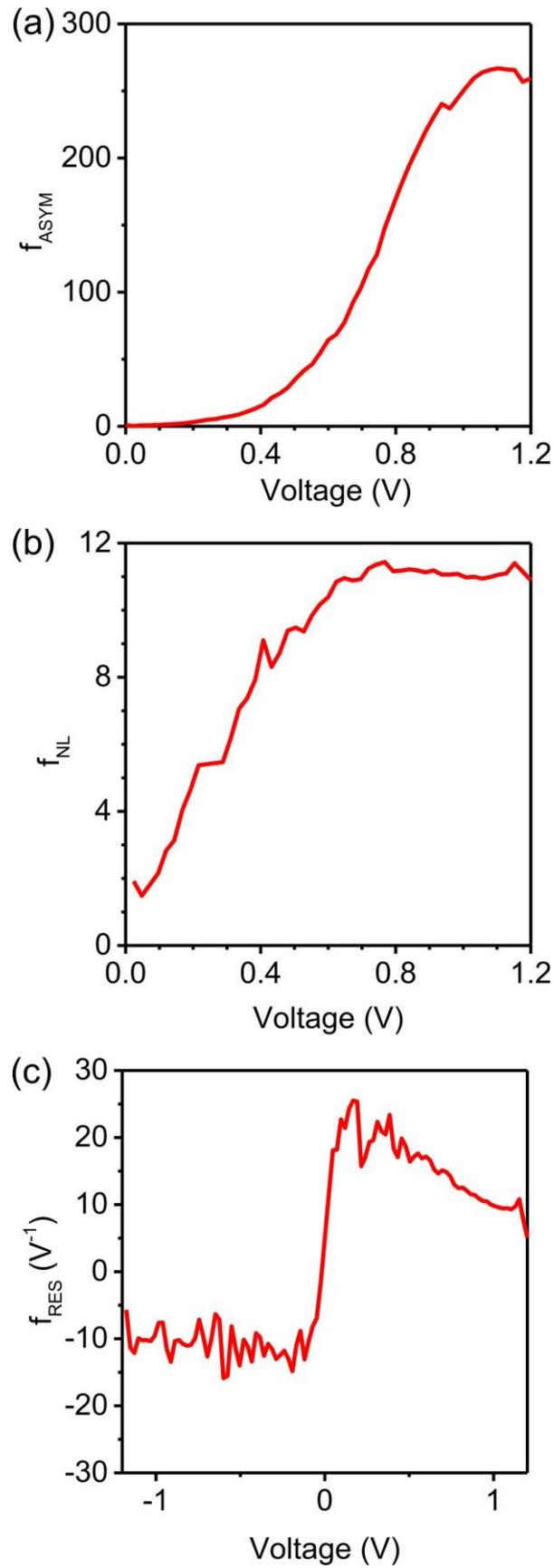

Figure 3. Typical FOMs of a MIG diode. (a) Asymmetry, (b) nonlinearity, and (c) responsivity as a function of applied bias voltage.

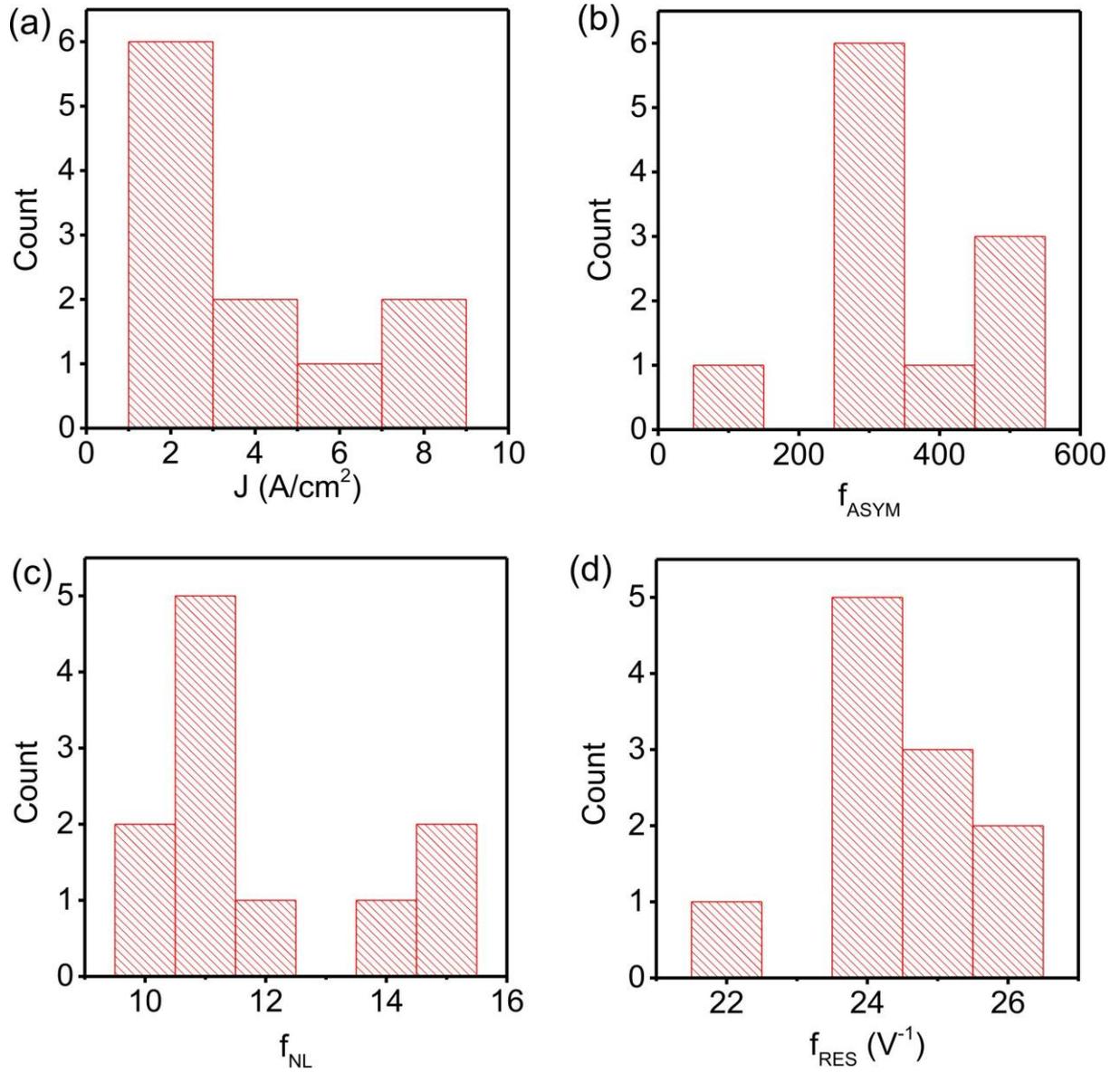

Figure 4. Histograms of the FOMs for 11 different MIG diodes on a single chip. (a) On-current density at 1 V bias. (b) Maximum asymmetry. (c) Maximum nonlinearity. (d) Maximum responsivity.

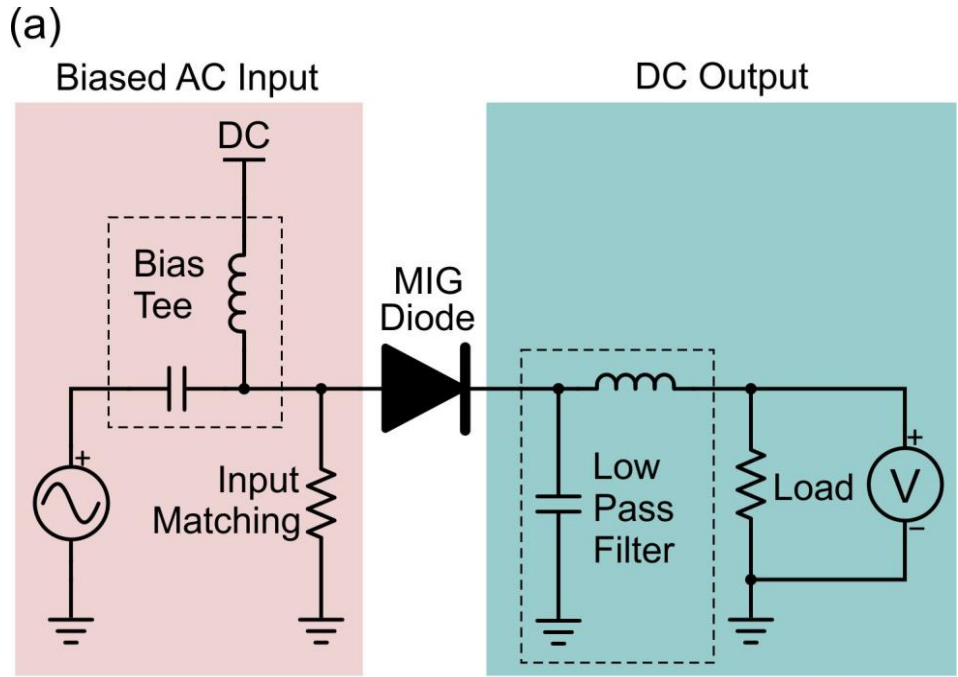

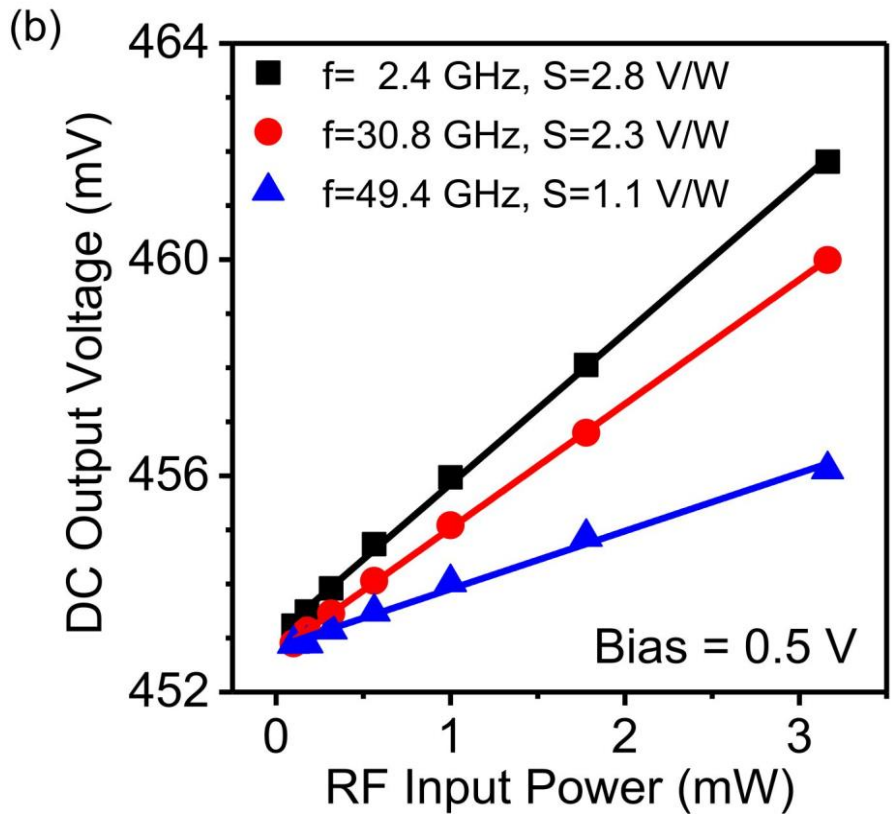

Figure 5. The high frequency power detection measurement of the MIG diode. (a) The setup for power detection measurement. (b) The DC output voltage versus RF input power at different frequencies. The corresponding high frequency power detection responsivity can be extracted from the slope of the linear relationship.

Table 1. Comparison of FOMs between MIM and MIG diodes. The on-current densities $J_{ON}$ are extracted at 1 V forward bias. The asymmetry, nonlinearity, and responsivity are the maximum values.

| **Stack Material** | $J_{ON}$(A/cm$^2$) | $f_{ASYM}$ | $f_{NL}$ | $f_{RES}$(V$^{-1}$) |
|---|---|---|---|---|
| Nb/Nb$_2$O$_5$(5nm)/Pt [9] | 2.0 | 9.8 | 8.2 | 16.9 |
| Nb/Nb$_2$O$_5$(15nm)/Pt [7] | N/A | 1500 | 4 | 20 |
| Ti/TiO$_2$/Bilayer Graphene [12] | 0.1 | 9000 | 8 | 10 |
| Ti/TiO$_2$/Graphene [Average this work] | 3.8 | 320 | 12 | 24 |
| Ti/TiO$_2$/Graphene [Max. value this work] | 7.5 | 520 | 15 | 26 |